\documentclass[showpacs,preprintnumbers,amsmath,amssymb,showkeys,aps,prd,a4paper,byrevtex]{revtex4}

\usepackage[dvips]{graphicx}
\usepackage{dcolumn}
\usepackage{bm}
\usepackage{epsfig}
\usepackage{amsmath}
\usepackage{amsfonts}
\usepackage{amssymb,amscd}
\usepackage[latin1]{inputenc}
\usepackage{units}
\usepackage{hyperref}
\def\pom{{I\!\!P}}

\begin{document}


\title{Probing the dilaton  in central exclusive processes at LHC}

\author{V. P. Goncalves}
\email{barros@ufpel.edu.br}
\author{W.~K. Sauter}
\email{werner.sauter@ufpel.edu.br}
\affiliation{High and Medium Energy Group, Instituto de F\'{\i}sica e Matem\'atica, Universidade Federal de Pelotas, Caixa Postal 354, CEP 96010-900, Pelotas, RS, Brazil}
\date{\today}

\begin{abstract}
The existence of a dilaton  as a pseudo-Nambu-Goldstone boson in spontaneous breaking of scale symmetry is predicted in beyond standard model theories  in which electroweak symmetry is broken via strongly coupled conformal dynamics. Such a particle is expected to have a mass below the conformal symmetry breaking scale $f$ and couplings to standard model particles similar to those of the SM Higgs boson.
In this paper  we estimate, for the first time, the dilaton production in exclusive processes considering Pomeron - Pomeron ($\pom \pom$) and photon - photon ($\gamma \gamma$) interactions, which are characterized by two rapidity gaps and intact hadrons in the final state. Our results indicate that if the dilaton is massive ($M_{\chi} \ge 2 M_W$), the study of dilaton production by $\pom \pom$ interactions in $pp$ collisions can be useful to determine its mass and the conformal energy scale. 

\end{abstract}


\maketitle


The prediction of the existence of new scalar particles is a characteristic of several candidate theories beyond the standard model (SM) (See e.g. Ref. \cite{review_plehn}). One of these particles is the dilaton, denoted as $\chi$, which is predicted to appear as a pseudo-Nambu-Goldstone boson in spontaneous breaking of scale symmetry \cite{Clark:1986gx}. In the particular scenario in which electroweak symmetry is broken via strongly coupled conformal dynamics, a neutral dilaton is expected with a mass below the conformal symmetry breaking scale $f$ and couplings to standard model particles similar to those of the SM Higgs boson. The searching of the dilaton in {\it inclusive} proton - proton collisions at LHC energies  motivated a lot of work, with special emphasis in the discrimination of the dilaton from the SM Higgs signals   \cite{Barger:2011nu,Coleppa:2011zx,Barger:2011hu,Coriano:2012nm,Chacko:2012vm,Bellazzini:2012vz,Jung:2014zga}.  
In particular, in Refs. \cite{Barger:2011nu,Coleppa:2011zx} the authors have derived regions of the mass $M_{\chi}$ and the conformal breaking scale  of the dilaton allowed by constraints from Higgs searches at LEP and LHC. They find that for low values of $f$, the dilaton is already excluded by the LHC in a large portion of parameter space ($M_{\chi} - f$) and that for large $f$ and large $M_{\chi}$ the dilaton is not excluded but could be discovered at LHC with more luminosity.
In this paper we extend these previous studies for {\it exclusive} processes, in which the hadrons colliding remain intact after the interaction, losing only a small fraction of their initial energy and escaping the central detectors \cite{forshaw_review}.  The signal would be a clear one with a dilaton tagged in the central region of the detector accompanied by regions of low hadronic activity, the so-called "rapidity gaps". In contrast to the inclusive production, which is characterized by large QCD activity and backgrounds which complicate the identification of a new physics signal, the exclusive production will be characterized by a clean topology associated to hadron - hadron interactions mediated by colorless exchanges. Our analysis is motivated by the studies performed in Refs.  \cite{outros1,Goncalves:2010dw,roman_tecni,thiel}, which demonstrated that central exclusive processes are very sensitive to Beyond Standard Model contributions.

\begin{figure}[t] 
\begin{center}
\includegraphics[scale=0.35] {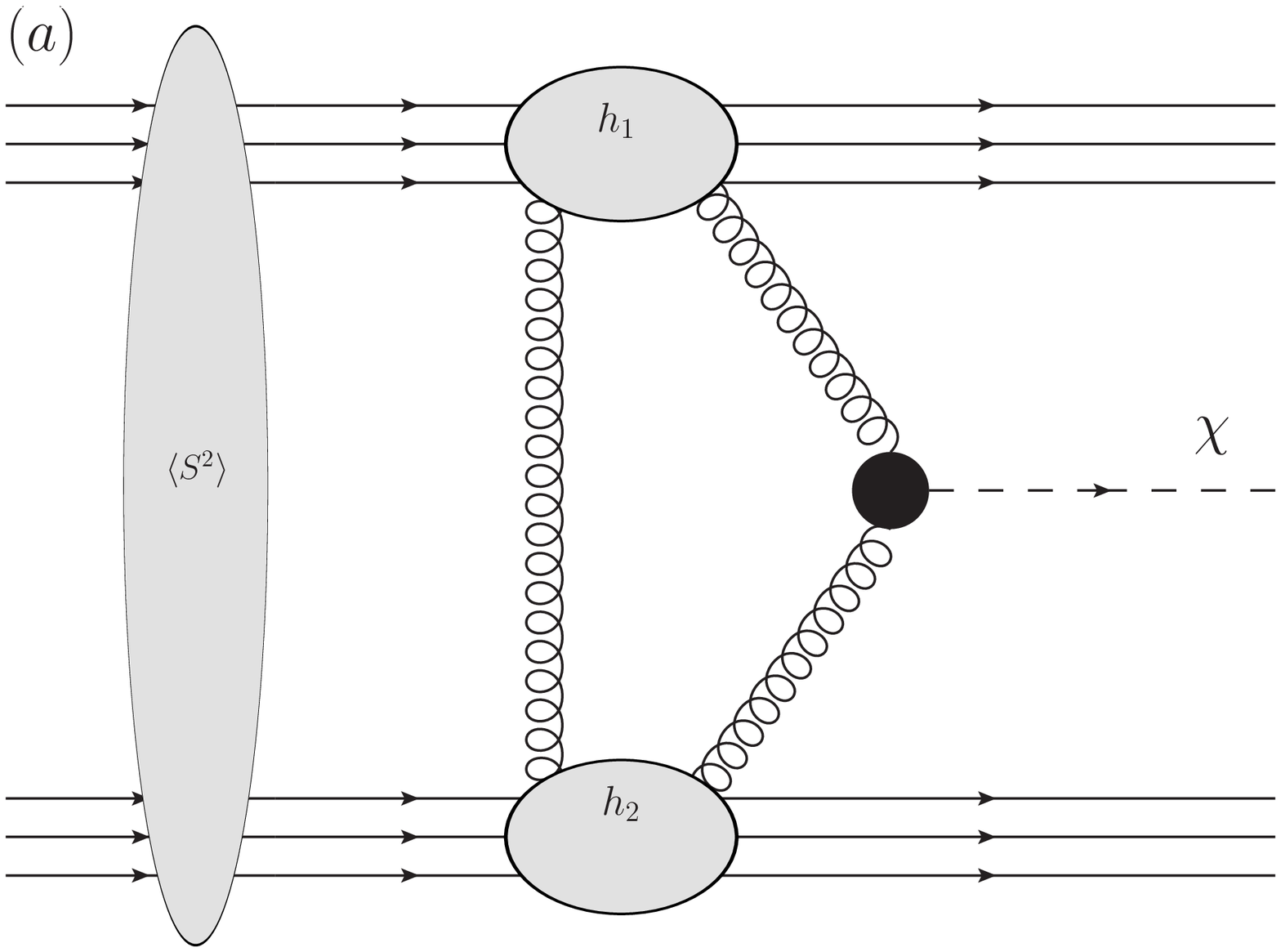} 
\includegraphics[scale=0.35] {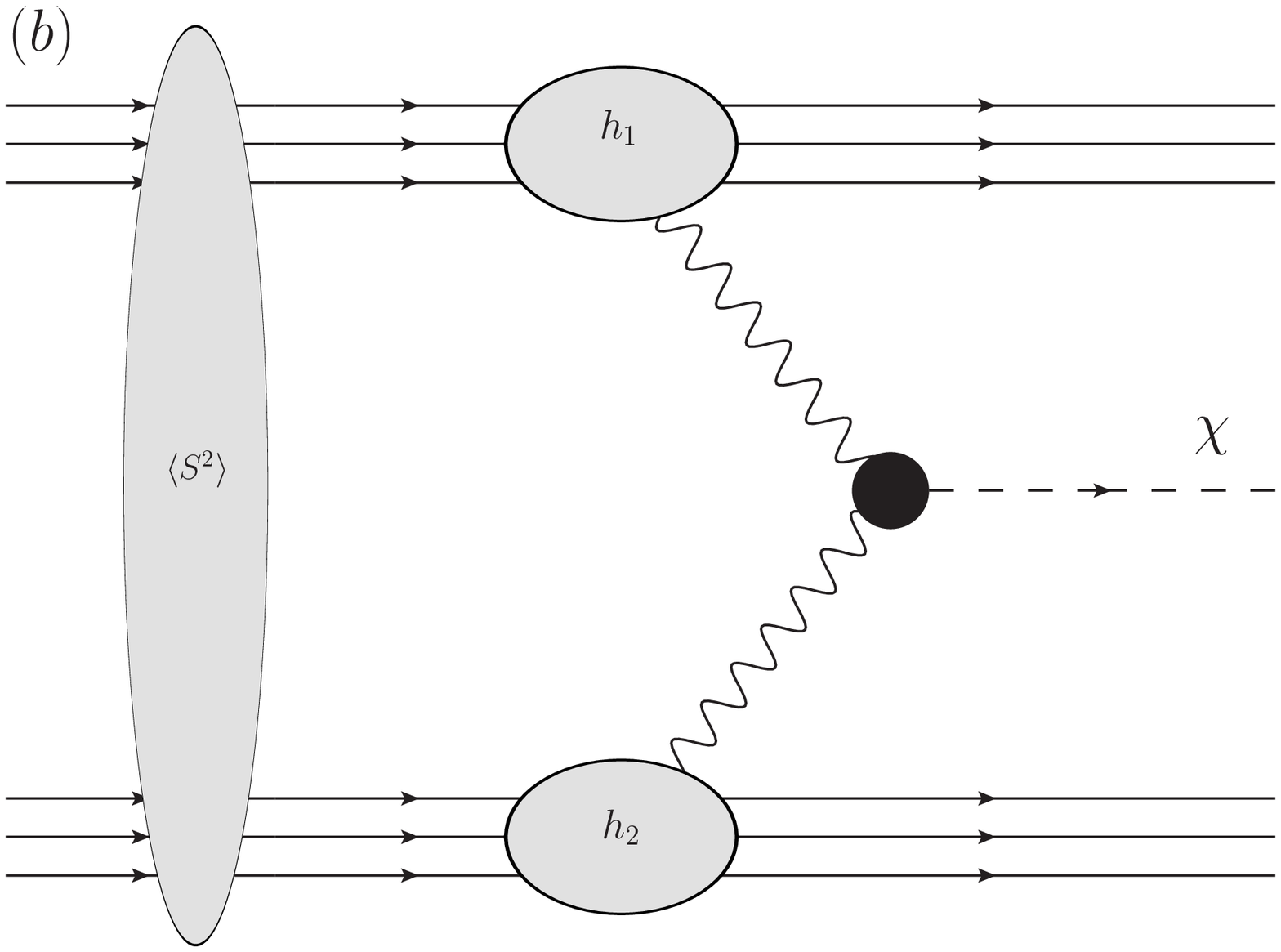}
\end{center}
\caption{Dilaton production in (a) Pomeron - Pomeron  and (b) photon - photon interactions. $\langle {\mathcal S}^2 \rangle$ is the gap survival probability, which gives the probability that secondaries, which are produced by soft rescatterings, do not populate the rapidity gaps.} \label{fig:1}
\end{figure}

In this letter we will calculate the dilaton production considering   Pomeron - Pomeron ($\pom \pom$) or photon - photon ($\gamma \gamma$) interactions in $pp$ and $PbPb$ collisions at LHC energies. These processes are represented 
in Figs.  \ref{fig:1} (a) and (b), respectively, and  can be written in the form
\begin{equation}
h_1 + h_2 \rightarrow h_1 \otimes {\chi} \otimes h_2,
\end{equation}
where $h_i$ is a proton or a nucleus  and  ${\chi}$ is the dilaton. The basic characteristic of these processes is the presence of two rapidity gaps ($\otimes$) in the final state, separating the dilaton from the intact outgoing hadrons. Experimentally, these processes have a very clear signal in the absence of pile-up, with the presence of the final state $\chi$ and no other hadronic activity seen in the central detector. Moreover, the measurement of the outgoing hadrons with installation of forward hadron spectrometers can be useful to separate the exclusive events \cite{forshaw_review}. Such possibility is currently under discussion in the ATLAS and CMS Collaborations at LHC. For comparison with our predictions for the dilaton production, we update previous estimates of the SM Higgs in exclusive processes and analyse the possibility of distinguish between these states in these processes.

In what follows we present a brief review of the theoretical description of  Pomeron - Pomeron ($\pom \pom$) or photon - photon ($\gamma \gamma$) interactions in $pp$ and $PbPb$ collisions at LHC energies. For the central exclusive production of a dilaton by pomeron - pomeron interactions we consider the model proposed by Khoze, Martin and Ryskin~\cite{Khoze:2000cy, Khoze:2001xm, Khoze:2000jm} some years ago, denoted Durham model hereafter, which has been used to estimate a large number of different final states and have  predictions in reasonable agreement with the observed rates for exclusive processes measured by the CDF collaboration~\cite{Aaltonen:2007am,Aaltonen:2007hs,Aaltonen:2009kg} and in the Run I of the LHC (For a recent review see Ref. \cite{review_lang}). In this model, the total cross section for central exclusive production  of a dilaton by $\pom \pom$ interactions  can be expressed in a factorizated form as follows
\begin{equation}
 \sigma = \int dy \langle {\mathcal S}^2 \rangle {\mathcal L}_{excl} \frac{2\pi^2}{M_{\chi}^3} \Gamma(\chi \rightarrow gg)\,\,, 
\label{eq:kmr}
\end{equation}
where $\langle {\mathcal S}^2 \rangle$ is the gap survival probability (see below),  $\Gamma$ stand for the partial decay width of the dilaton $\chi$ in a pair of gluons   and ${\mathcal L}_{excl}$ is the effective luminosity, given by
\begin{equation}
 {\mathcal L}_{excl} = \left[ {\cal{C}} \int \frac{dQ_t^2}{Q_t^4} f_g(x_1,x_1^{\prime},Q_t^2, \mu^2) f_g(x_2,x_2^{\prime},Q_t^2, \mu^2)\right]^2\,\,,
\end{equation}
where ${\cal{C}} = \pi/[(N_c^2 - 1)b]$, with $b$ the $t$-slope ($b = 4$ GeV$^{-2}$ in what follows), $Q_t^2$ is the virtuality of the soft gluon needed for color  screening, $x_1$ and $x_2$ being  the longitudinal momentum of the gluons which participate of the hard subprocess  and $x_1^{\prime}$ and $x_2^{\prime}$ the longitudinal momenta of the spectator gluon. Moreover, the quantities $f_g$ are the  skewed unintegrated gluon densities. Since 
\begin{eqnarray}
(x^{\prime} \approx \frac{Q_t}{\sqrt{s}}) \ll (x \approx \frac{M_{\chi}}{\sqrt{s}}) \ll 1
\end{eqnarray}
 it is possible to express $f_g(x,x^{\prime},Q_t^2, \mu^2)$, to single log accuracy, in terms of the conventional integrated gluon density $g(x)$, together with a known Sudakov suppression $T$ which ensures that the active gluons do not radiate in the evolution from $Q_t$ up to the hard scale $\mu \approx M_{\chi}/2$.  Following  \cite{Khoze:2001xm} we will assume that
\begin{equation}
f_g(x,x^{\prime},Q_t^2, \mu^2) = S_g \frac{\partial}{\partial \ln Q_t^2} \left[ \sqrt{T(Q_t,\mu)}\ xg(x,Q_t^2) \right]\,\,,
\end{equation}
where  $S_g$ accounts for the single $\log Q^2$ skewed effect, being $S_g \sim 1.2 (1.4)$ for LHC (Tevatron) (For a more detailed discussion about $S_g$ see Ref. \cite{harland}). The Sudakov factor $T(Q_t,\mu)$ is given by
\begin{eqnarray}
T(Q_t,\mu)  =  \exp \left\{ -\int_{Q_t^2}^{\mu^2} \frac{dk_t^2}{k_t^2} \frac{\alpha_s(k_t^2)}{2\pi} \int_{0}^{1-\Delta} dz \, \left[ zP_{gg}(z) + \sum_{q} P_{qg}(z) \right] \right\},
\end{eqnarray} 
with $k_t$ being an intermediate scale between $Q_t$ and $\mu$, $\Delta = k_t/(\mu + k_t)$, and $P_{gg}(z)$ and $P_{qg}(z)$ are the leading order Dokshitzer - Gribov - Lipatov - Altarelli - Parisi (DGLAP) splitting functions \cite{dglap}. 
In this paper we will calculate  $f_g$ in the proton case considering that  the integrated gluon distribution $xg(x,Q_T^2)$ is described by the MSTW parametrization \cite{mstw}. In the nuclear case we will include the shadowing effects in $f_g^A$ considering that the nuclear gluon distribution is given by the  EPS09 parametrization~\cite{Eskola:2009uj}, where  
\begin{eqnarray}
 x g_A(x,Q_T^2) = A R_g^A(x, Q_T^2) x g_p(x, Q_T^2)\,\,,
\end{eqnarray}
with $R_g^A$ describing the nuclear  effects in $xg_A$ and $A$ the number of mass of the nucleus. 
Moreover, the partial decay width of the dilaton into two gluons is given by \cite{Coleppa:2011zx}
\begin{equation} \label{eq:lamgggg}
\Gamma_{{\chi} \rightarrow gg}(M_{\chi}) = {\cal{C}}_{g} \frac{\upsilon^2}{f^2} \frac{G_F \alpha_s^2 M_{\chi}^3}{36\sqrt{2} \pi^3}\left| \frac{3}{4} \sum_f F_{1/2}(\tau_f) \right|^2,
\end{equation}
where  $\upsilon = \unit[246]{GeV}$ is the scale of electroweak symmetry breaking, $f$ is the energy scale of conformal scale (see discussion below), 
$G_F$ is the Fermi constant and $\alpha_s$ is the strong running coupling. The scaling variables are $\tau_f = 4m_f^2/M_{\chi}^2$, $\tau_W = 4m_W^2/M_{\chi}^2$ and the sums runs over all fermions. The loop functions are given by the following expressions\cite{Gunion:1989we},
\begin{equation}
 F_1(\tau) = 2 + 3\tau + 3\tau(2 -\tau)f(\tau)
\end{equation}
\begin{equation}
 F_{1/2}(\tau) = -2\tau\left[ 1 + (1-\tau)f(\tau) \right]
\end{equation}
where 
\begin{equation}
 f(\tau) = \left\lbrace
\begin{array}{l@{\ ,\ }l}
\left[ \sin^{-1}(1/\sqrt{\tau}) \right]^2 & \tau \geq 1 \\
-\frac{1}{4}\left[ \ln\left(\frac{1 + \sqrt{1-\tau}}{1 - \sqrt{1-\tau}} \right) - i\pi \right]^2 & \tau < 1
\end{array} \right.
\end{equation}
with $\tau = 4m_i^2/M_{\chi}^2$.
Moreover,  the coefficient  ${\cal{C}}_{g}$ in Eq. (\ref{eq:lamgggg}) is given by
\begin{equation} \label{eq:cg}
{\cal{C}}_{g} = \frac{|-b_G + 1/2\sum_{i=q} F_{1/2}(\tau_i)|^2}{|1/2\sum_{i=f} F_{1/2}(\tau^2)|^2}\,\,, 
\end{equation}
where $b_G = 11 - \frac{2}{3}n_f$, with $n_f = 6$, and  the  sum runs over quarks only.

On the other hand, the cross section for the exclusive dilaton production in the two-photon fusion  process, Fig. \ref{fig:1} (b),  is given by \cite{baur_jpg}
\begin{eqnarray}
\sigma = \langle {\mathcal S}^2 \rangle \int_{0}^{\infty}\! \frac{d\omega_{1}}{\omega_{1}}\! \int_{0}^{\infty}\! \frac{d\omega_{2}}{\omega_{2}}\ F(\omega_1, \omega_{2})\ \hat{\sigma}_{\gamma \gamma \rightarrow {\chi}}(\omega_{1}, \omega_{2})
\end{eqnarray}
where $\hat{\sigma}_{\gamma \gamma \rightarrow {\chi}}$ is the cross section for the subprocess $\gamma \gamma \rightarrow {\chi}$, $\omega_1$ and $\omega_2$ the energy of the photons which participate of the hard process and  $F$ is the folded spectra of the incoming particles (which corresponds to an ``effective luminosity'' of photons) which we assume to be given by~\cite{baur_ferreira}
\begin{equation}
 F(\omega_1, \omega_{2}) = 2\pi \int_{R_{A}}^{\infty} db_{1} b_{1} \int_{R_{B}}^{\infty} db_{2} b_{2} \int_{0}^{2\pi} d\phi\ N_{1}(\omega_{1}, b_{1}) N_{2}(\omega_{2}, b_{2}) \Theta(b - R_{A} - R_{B}) \label{efe}
\end{equation}
where $b_i$ is the impact parameter of the hadrons in relation to the photon interaction point, $\phi$ is the angle between $\mathbf{b}_1$ and $\mathbf{b}_2$, $R_i$ are the projectile radii  and $b^2 = b_1^2 + b_2^2 - 2b_1 b_ 2 \cos \theta$. 
The theta function in Eq. (\ref{efe}) ensures that the hadrons do not overlap \cite{baur_ferreira}. The Weizs\"acker-Williams photon spectrum for a given impact parameter is given in terms of the nuclear charge form factor $F(k_{\perp}^2)$, where $k_{\perp}$ is the four-momentum of the quasi-real photon, as follows  \cite{upcs}
\begin{eqnarray}
N(\omega,b) = \frac{\alpha Z^2}{\pi^2 \omega}\left| \int_0^{+\infty}  dk_{\perp} k_{\perp}^2 \frac{F\left((\frac{\omega}{\gamma})^2 + \vec{b}^2\right)}{(\frac{\omega}{\gamma})^2 + \vec{b}^2} \cdot J_1(b k_{\perp}) \right|^2 \,\,,
\end{eqnarray}
where $J_1$  is the Bessel function of the first kind. For a point-like nucleus one obtains that \cite{upcs}
\begin{equation}
N(\omega,b) =\frac{\alpha_{em} Z^2}{\pi^2} \left(\frac{\xi}{b}\right)^2 \left\{ K_1^2(\xi) + \frac{1}{\gamma^2}  K^2_0(\xi)\right\} ,
\label{ene}
\end{equation}
with $K_{0,1}$ being the modified Bessel function of second kind, $\xi = \omega b/\gamma v$, $v$  the velocity of the hadron, $\gamma$  the Lorentz factor and $\alpha_{em}$  the electromagnetic coupling constant. This expression have been derived considering a semiclassical description of the electromagnetic interactions in peripheral collisions, which works very well for heavy ions (See e.g. \cite{baur_jpg}).  For  protons, it is more appropriate to obtain the equivalent photon spectrum  from its elastic form factors in the dipole approximation (See e.g. \cite{david}). An alternative  is to use Eq. (\ref{ene}) assuming $R_p = 0.7$ fm for the proton radius, which implies a  good agreement with the parametrization of the luminosity obtained in \cite{Ohnemus:1993qw} for proton-proton collisions. We will assume this procedure in what follows. The $\gamma \gamma \rightarrow \chi$ cross section can be expressed as follows
\begin{equation} \label{eq:epa}
\hat{\sigma}_{\gamma \gamma \rightarrow {\chi}}(\omega_{1}, \omega_{2}) = \int ds\ \delta(4\omega_{1}\omega_{2} -s) \frac{8\pi^2}{M_{\chi}} \Gamma_{{\chi} \rightarrow \gamma \gamma}(M_{\chi}) \delta(s - M_{\chi}^2),
\end{equation}
where partial decay width of the dilaton into two photons,  $\Gamma_{{\chi} \rightarrow \gamma \gamma}$, was calculated in \cite{Coleppa:2011zx} and is given by:
\begin{equation} \label{eq:lamgamgam}
\Gamma_{{\chi}\rightarrow \gamma\gamma}(M_{\chi}) ={\cal{C}}_{ \gamma } \frac{\upsilon^2}{f^2} \frac{G_F \alpha_\mathrm{em} M_{\chi}^3}{128 \sqrt{2} \pi^3} \left| F_1(\tau_W) + \sum_f N_c Q_f^2 F_{1/2}(\tau_f) \right|^2
\end{equation}
where $\alpha_\mathrm{em}$ is the electromagnetic coupling constant, $N_c$ is the number of colors, $Q_f$ is the fermion charge. Moreover, the coefficient ${\cal{C}}_{ \gamma }$ is given by \cite{Coleppa:2011zx}
\begin{eqnarray}
R_\gamma = \frac{|-b_{EM} + \sum_{i=f,b}N_{c,i} Q_i^2 F_i(\tau_i)|^2}{|\sum_{i=f,b}N_{c,i} Q_i^2 F_i(\tau_i)|^2}
\end{eqnarray}
where $b_{EM} = - 11/3$, the sum runs over fermions ($f$) and bosons ($b$), $N_{c,i}$ is the color multiplicity number ($N_{c,i}=1$ for  bosons and leptons and $N_{c,i}=3$ for quarks) and $Q$ is the electric charge in units of $e$.

\begin{figure}[t]
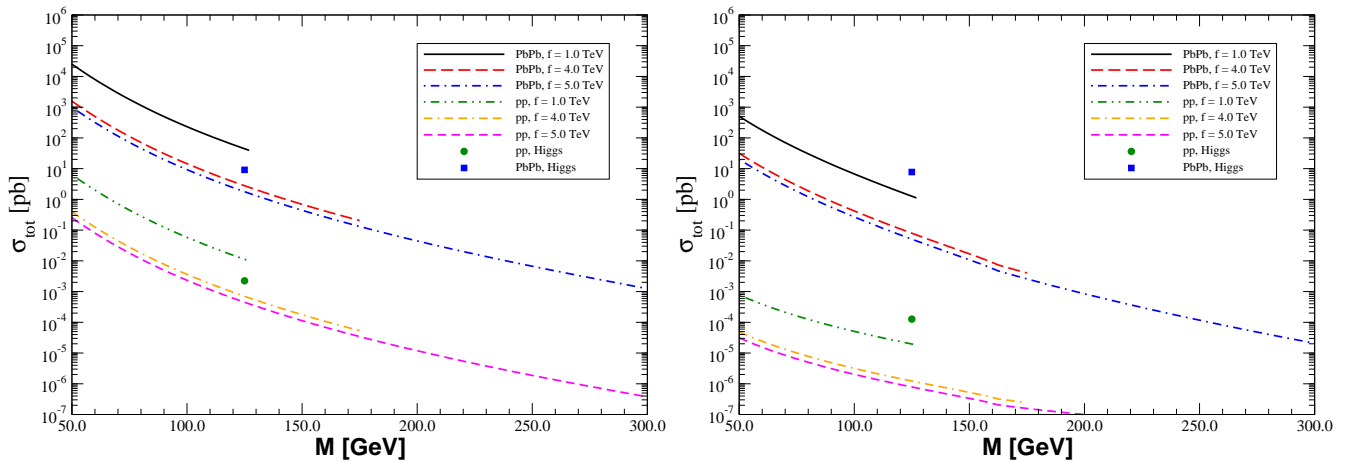
 
\begin{center}
\includegraphics[scale=0.35] {dilaton_gluon4.eps}  
\includegraphics[scale=0.35] {dilaton_photon4.eps}
\end{center}
\caption{(Color online) Total cross section for the dilaton production in (a) Pomeron - Pomeron and (b) photon - photon interactions in $pp$ and $PbPb$ collisions as a function of the dilaton mass. The corresponding predictions for the SM Higgs production are also presented for comparison.  } \label{fig:2}
\end{figure} 

In what follows we will present our predictions for the dilaton production in $pp$ ($\sqrt{s} = 14$ TeV) and $PbPb$ ($\sqrt{s} = 5.5$ TeV) collisions at LHC energies. We  assume $m_W = \unit[80.4]{GeV}$, $m_t=\unit[173]{GeV}$, $m_b = \unit[4.2]{GeV}$, $m_c=\unit[1.4]{GeV}$ and $m_\tau = \unit[1.77]{GeV}$. Moreover, in order to obtain realistic predictions for the exclusive production of the dilaton, it is crucial to use an adequate value for the gap survival probability, $\langle {\mathcal S}^2 \rangle$. This factor is the probability that secondaries, which are produced by soft rescatterings do not populate the rapidity gaps, and depends on the particles involved in the process and  in the center-of-mass energy.  
For the case of the central exclusive production described by the Durham model we will assume that $\langle {\mathcal S}^2 \rangle = 3 \, \%$ for proton - proton collisions at LHC  energies~\cite{Khoze:2001xm}. However, the value of the corresponding value of the  survival probability for nuclear collisions still is an open question. Here we assume the  conservative estimate proposed in Ref. \cite{Goncalves:2010dw} which assume that :  $\langle {\mathcal S}^2 \rangle_{A_1A_2} = \langle {\mathcal S}^2 \rangle_{pp}/(A_1 \cdot A_2)$. In contrast, for  two-photon interactions, it is expected that the contribution of secondary interactions for the cross section will be negligible \cite{forshaw_review}. For simplicity, we will assume that  $\langle {\mathcal S}^2 \rangle_{A_1A_2} = \langle {\mathcal S}^2 \rangle_{pp} = 1$ for the dilaton production by two photons. However, this subject deserves a more detailed analysis. For our calculations of the SM Higgs production in exclusive processes, we will assume the same parameters described before and will take   ${\cal{C}}_{ \gamma/g } \frac{\upsilon^2}{f^2} = 1$ in Eqs. (\ref{eq:lamgggg}) and (\ref{eq:lamgamgam}). Moreover, we assume $M_H = 125$ GeV.

\begin{figure}[t]
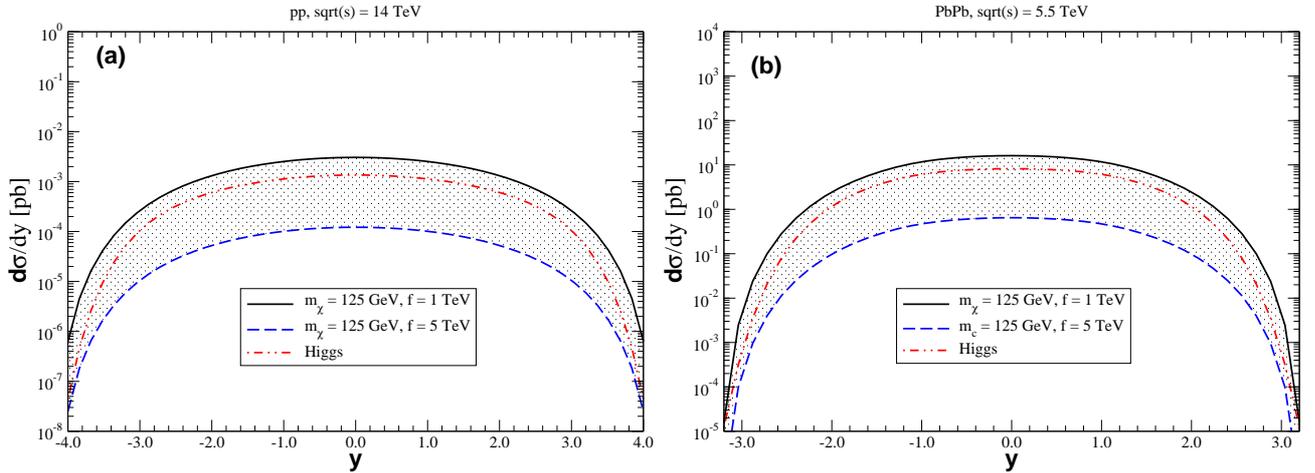
 
\begin{center} 
\includegraphics[scale=0.35] {dilaton_dsdyy3.eps} 
\includegraphics[scale=0.35] {dilaton_dsdyyN3.eps}
\end{center}
\caption{(Color online) Rapidity distribution for the dilaton production in $\pom \pom$ interactions considering (a) $pp$ and (b) $PbPb$ collisions at LHC energies. The corresponding predictions for the SM Higgs production are also presented for comparison.  } \label{fig:3}
\end{figure} 

In Fig. \ref{fig:2} (a) and (b) we present our results for the dependence of the total cross sections on the mass of the dilaton considering $\pom \pom$ and $\gamma \gamma$ interactions in $pp$ and $PbPb$ collisions, respectively. For comparison, we also present the corresponding predictions for the SM Higgs production. Following previous studies
\cite{Barger:2011hu, Coleppa:2011zx}, which have derived allowed regions 
for the dilaton mass $M_{{\chi}}$ and  conformal energy scale  $f$ by considering the LHC data relevant to the double gauge boson decays of the dilaton, we restrict our analysis for the following ranges: (a) $f = 1.0$ TeV and $M_{{\chi}} \le 126$ GeV; (b) $f = 4.0$ TeV and $M_{{\chi}} \le 175$ GeV and (c) $f = 5.0$ TeV, with all  values of $M_ {{\chi}}$ being allowed \cite{Barger:2011hu, Coleppa:2011zx}. As the partial decay widths of the dilaton into gluons and photons, Eqs. (\ref{eq:lamgggg}) and (\ref{eq:lamgamgam}),  are proportional to $1/f^2$, we obtain that the cross sections are strongly dependent on the choice of $f$, decreasing at larger values of the conformal energy scale.
 For the case of $\pom \pom$ interactions, Fig. \ref{fig:2} (a), we obtain that the predictions for the dilaton production in $PbPb$ collisions is three orders of magnitude larger than for $pp$ collisions.  In comparison with the SM Higgs predictions, we obtain that for $M_{\chi} = 125$ GeV   the dilaton cross section is larger than the exclusive SM Higgs cross section for $f = 1.0$ TeV and a factor $\ge$ 2 smaller for $f \ge 4.0$ TeV.	   In contrast, for $\gamma \gamma$ interactions, Fig. \ref{fig:2} (b), we obtain that at small $M_{\chi}$ the predictions for the dilaton production in $PbPb$ collisions is six orders of magnitude larger than for $pp$ collisions, which is directly associated to the $Z^2$ dependence of the nuclear photon flux. Moreover, we obtain that this difference decreases at larger values of $M_{\chi}$, which is associated to the fact that the  maximum value of the photon energy in the photon flux is given by $\gamma/b_{max}$ \cite{upcs}, where $\gamma$ is the Lorentz factor and $b_{max} $ is proportional to the  hadron radius. Consequently, the proton photon flux contains a larger number of energetic photons, 
which increases the cross section for the production of a massive final state in $pp$ collisions  
 in comparison to the nuclear case.  However,  we predict very small values for  the dilaton production induced by two-photon fusion in $pp$ cross section. Finally, in comparison to the SM Higgs cross section for $\gamma \gamma$ interactions, we obtain for $M_{\chi} = 125$ GeV and $f= 1.0$ TeV that the dilaton cross section is a factor five smaller than the SM Higgs one.  

In Fig. \ref{fig:3} (a) and (b) we present our results for the rapidity distribution for the dilaton production in $\pom \pom$ interactions considering  $pp$ and $PbPb$ collisions at LHC energies, respectively. In these figures we assume $M_{\chi} = M_H = 125$ GeV. In agreement with the results presented in Fig. \ref{fig:2} (a), we obtain that the predictions are strongly dependent on $f$ and the dilaton production in nuclear collisions is four orders of magnitude than in $pp$ collisions.  
For  $f = 1$ TeV, we predict larger values of the rapidity distribution for the dilaton production than the SM Higgs one.

\begin{table}[t]
\begin{center}
\begin{tabular} {||c|c|c||}
\hline
\hline
{\bf $\pom \pom$ interactions} & {\bf $pp$ collisions}  & {\bf $PbPb$ collisions} \\
\hline
\hline
Dilaton & $1100$ ($40$) & $21.1\times 10^{-3}$ ($7.35\times 10^{-4}$) \\
\hline
Higgs & $200$ & $5\times 10^{-3}$ \\
\hline
\hline
{\bf $\gamma \gamma$ interactions} & {\bf $pp$ collisions}  & {\bf $PbPb$ collisions} \\
\hline
\hline
Dilaton & $1.9$ ($7.3\times 10^{-2}$) &  $0.6 \times 10^{-3}$ ($2.5\times 10^{-5}$) \\
\hline
Higgs & $19$ & $4.9\times 10^{-3}$ \\
\hline
\hline
\end{tabular}
\end{center}
\caption{Number of events by year for the production of a  dilaton in $pp$ and $PbPb$ collisions considering $\pom \pom$ and $\gamma \gamma$ interactions. Values obtained for $M_{\chi} = 125$ GeV and $f = 1.0 \, (5.0)$ TeV. The corresponding predictions for the SM Higgs production are also presented for comparison. }
\label{tab:1}
\end{table}

In Table \ref{tab:1} we present our results for  the production rates for dilaton production at LHC energies  considering the $\pom \pom$ and $\gamma \gamma$ interactions assuming that $M_{\chi} = M_H = 125$ GeV.  
 At LHC we assume the  design luminosities ${\cal L} = 10^7  /\, 0.5$ mb$^{-1}$s$^{-1}$ for $pp/PbPb$  collisions at $\sqrt{s} = 14/\,5.5$ TeV and a run time of $10^7 \, (10^6)$ s for collisions with protons (ions). The predictions for the SM Higgs production also are presented for comparison.  Due to the small luminosity for heavy ion collisions, we obtain very small values for the event rates of Higgs and dilaton production in $\pom \pom$ and $\gamma \gamma$ interactions. In contrast, we obtain that the event rates  are a factor $\ge 10^{3}$ larger than for $pp$ collisions. For $\gamma \gamma$ interactions in $pp$ collisions we obtain that the predictions for the dilaton production ($f=1.0$ TeV) are one order of magnitude smaller than the SM Higgs one, as expected from the Fig. \ref{fig:2}. Moreover, the predictions for the dilaton and SM Higgs production in $\pom \pom$ interactions are a factor $\ge 10$ larger than those for $\gamma \gamma$ interactions. In particular, we obtain that the dilaton production for $f=1.0$ TeV is a factor five  larger than the SM Higgs one due to  the magnitude of the factor $C_g v^2 / f^2$ in Eq. (\ref{eq:lamgggg}), which is larger of one for this value of the conformal energy scale. At larger values of $f$, the Higgs production dominates due to the $1/f^2$ dependence of the dilaton cross section.

Lets discuss now the potential backgrounds for the dilaton production in exclusive processes.
The partial widths for the dilaton to decay into any SM final state  were estimated in Refs. \cite{Barger:2011hu, Coleppa:2011zx}. Depending on the dilaton mass,  different channels seem more favourable, and each of the decay channels has its own difficulties for the experimental identification. 
At $M_{\chi} < 2 M_W$ the dilaton decay is characterized by a large $gg$ branching fraction, in contrast to the Higgs decay which is dominated by the decay in $b \bar{b}$ pair for $M_H \le 140$ GeV. This distinct behaviour is due to the enhancement of the $\chi gg$ coupling via the QCD beta function coefficient in Eq. (\ref{eq:cg}). 
In contrast, for  $M_{\chi} > 2 M_W$ the dilaton decays predominantly to $WW$, $ZZ$ and $t\bar{t}$.
Consequently, the main backgrounds are the $gg$ and $WW$ production in exclusive processes. The cross sections for these processes were estimated  e.g. in Refs. \cite{roman1,roman2}. As demonstrated in Ref. \cite{roman1}, the $gg$ contribution dominates the central exclusive production of dijets.
Comparing our results with those obtained in Ref. \cite{roman1}, we obtain that our predictions for the production of a dilaton with mass $M_{\chi} < 140$ GeV  which decays into a $gg$ final state is  two orders of magnitude smaller than the exclusive $gg$ dijet production. Consequently, the identification of a light dilaton in exclusive processes considering this final state will be not possible.  On the other hand, for the production of a massive dilaton which decays into a $WW$ pair, we need to compare our predictions with those presented in Ref. \cite{roman2}, which have demonstrated that the central exclusive $W^+W^-$ production is dominated by the two - photon fusion.  We obtain that also in this case the signal of the dilaton production will be smaller than the background, in particular for large $M_{\chi}$ masses, since the cross section for the $pp \rightarrow pp W^+W^-$ (via $\gamma \gamma$) process is weakly dependent on the  invariant mass $M_{WW}$ (See Fig. 14 in Ref. \cite{roman2}). However, if  the $\gamma \gamma$ and $\pom \pom$ processes are separated by measuring the four-momentum transfers squared in the proton lines, as planned for future studies at ATLAS and CMS, we have that the signal will be similar to the background associated to the $pp \rightarrow pp W^+W^-$ (via $gg$) process. An alternative to search a massive dilaton in exclusive processes, is to consider its decay into a $ZZ$ pair and/or a Higgs pair. As demonstrated in Ref. \cite{Barger:2011hu}, these two decay channels are similar to the $WW$ one for $M_{\chi}> 200$ GeV. The exclusive $ZZ$ production has been estimated in Ref. \cite{thiel} as a probe of large extra dimension scenario. Comparing our results with the SM predictions  presented in \cite{thiel}, we obtain that the signal will be larger than the background for large invariant masses of this final state. The magnitude of the  double Higgs production in exclusive processes still is an open question, but it is expected to be very small. Consequently, the search of the dilaton considering its decay into a Higgs pair can also be a promising way.

Finally, lets summarize our main results and conclusions. In this paper  we estimated, for the first time, the dilaton production in exclusive processes considering $\pom \pom$ and $\gamma \gamma$ interactions, which are characterized by two rapidity gaps and intact hadrons in the final state. Our goal was to verify if exclusive processes can be considered a viable alternative to the proposed searches of the dilaton in inclusive processes. In contrast to inclusive processes, where the incident hadrons dissociate and the final state is populated by a large number of particles, which makes the separation of the dilaton a hard task, in exclusive processes the incident hadron remains intact and the dilaton will be centrally produced, separated from the very forward hadrons by large rapidity gaps.
Consequently,  in exclusive processes the dilaton is expected to be produced in a clean environment.
Moreover, if the momenta of the outgoing hadrons are measured by forward detectors, the mass of the dilaton is expected to be reconstructed with very precise resolution. All these aspects have motivated the  analysis performed in this letter. We have considered one of the possible scenarios which predicts the dilaton and could be analysed at LHC,  in which the scale invariance of the strong dynamics is manifest at very high energy but is spontaneously broken at a scale $f$, not too above the electroweak scale.  Consequently, the identification of the  dilaton provide a hint to the conformal nature of the strong sector.  Taking into account the constraints in $f$ and $M_{\chi}$ from the Higgs searches at LEP and LHC, we have estimated the dilaton cross sections and event rates for the dilaton production induced by   $\pom \pom$ and $\gamma \gamma$ interactions in $pp$ and $PbPb$ collisions. For a dilaton with mass identical to the SM Higgs, we predict   larger cross sections in comparison to the SM Higgs production in Pomeron - Pomeron interactions for  $f = 1.0$ TeV, which is the minimal value of $f$ allowed for $M_{\chi} = 125$ GeV. At larger values  of $f$, the dilaton cross section becomes smaller than the Higgs one due to its $1/f^2$ dependence. In contrast, for photon - photon interactions, the dilaton cross section is smaller to the SM Higgs one for all allowed values of $f$. Taking into account the main decay channels of the dilaton, we have compared our predictions with potential backgrounds. Our results demonstrated that for a light dilaton the signal-to-background ratio will be very small, which implies that the probe of a light dilaton in exclusive processes at LHC will be not possible. In contrast, our results indicated that for a massive dilaton, which can be produced if $f > 4.0$ TeV, the signal-to-background ratio will be favourable if the  dilaton decay channels into $ZZ$ and/or $HH$ are considered. Our main conclusion is that the study of exclusive processes can be useful to search a massive dilaton, as well as  to constrain the  mass and the conformal energy scale.

\section*{Acknowledgements}
This work was partially financed by the Brazilian funding agencies FAPERGS, CNPq and CAPES.


\end{document}